\documentclass[dvipdfmx]{ptephy}

\preprintnumber{KIAS-P13068\\ KEK-TH-1696} 

\usepackage{epsfig,multicol,multirow}
\usepackage{amsmath,latexsym,amssymb}
\usepackage{here}

\numberwithin{equation}{section}

\usepackage{cite}

\newcommand\fverb{\setbox\fverbbox=\hbox\bgroup\verb}
\newcommand\fverbdo{\egroup\medskip\noindent%
			\fbox{\unhbox\fverbbox}\ }
\newcommand\fverbit{\egroup\item[\fbox{\unhbox\fverbbox}]}
\newbox\fverbbox

\newcommand {\beq} {\begin{equation}}
\newcommand {\eeq} {\end{equation}}
\newcommand {\beqa}{\begin{eqnarray}}
\newcommand {\eeqa}{\end{eqnarray}}

\newcommand {\tr}{{\rm tr\,}}
\newcommand {\Tr}{\mbox{Tr\,}}

\newcommand {\Pf}{\mbox{Pf}}

\begin{document}

\title{A renormalization group method for studying 
the early universe in the Lorentzian IIB matrix model}

\author{\name{Yuta Ito}{1,\ast}, \name{Sang-Woo Kim}{2,\ast}, \name{Yuki Koizuka}{1,\ast}, \name{Jun Nishimura}{1,3,\ast} and \name{Asato Tsuchiya}{4,\ast}}

\address{\affil{1}{Department of Particle and Nuclear Physics,\\
Graduate University for Advanced Studies (SOKENDAI),\\
Tsukuba, Ibaraki 305-0801, Japan}
\affil{2}{School of Physics, Korea Institute for Advanced Study (KIAS),\\
85 Hoegiro Dongdaemun-gu, Seoul 130-722, Korea\\
}
\affil{3}{Theory Center, High Energy Accelerator Research Organization (KEK),\\
Tsukuba, Ibaraki 305-0801, Japan}
\affil{4}{Department of Physics, Shizuoka University,\\
836 Ohya, Suruga-ku, Shizuoka 422-8529, Japan}
\email{yito@post.kek.jp,sang@kias.re.kr,%
koizuka@post.kek.jp,
jnishi@post.kek.jp,satsuch@ipc.shizuoka.ac.jp}}

\begin{abstract}%
We propose a new method for studying
the early universe in the Lorentzian version of the IIB matrix model,
which is considered to be 
a nonperturbative formulation of superstring theory.
This method is based on the idea of renormalization group,
and it enables us 
to study the time-evolution of the universe for much longer time than
in the previous work, 
which showed that the SO(9) rotational symmetry 
is spontaneously broken down to SO(3)
after a ``critical time''.
We demonstrate how this method works in a simplified model, which
is expected to capture the behaviors of the original model
when the space is not so large.
In particular, we present clear evidence that 
the three-dimensional space
expands \emph{exponentially}
after the critical time
in this simplified model.
\end{abstract}

\subjectindex{B25}

\maketitle

\section{Introduction}
\label{sec:introduction}

Understanding how our universe began is one of the most fundamental
and fascinating themes in theoretical physics.
For instance, there are good reasons to believe
that our universe underwent a rapid expansion called inflation
before the Big Bang.
While there are various models which describe
inflation phenomenologically, 
we still do not have a description 
based on first-principle calculations in 
a complete quantum gravity theory like superstring theory.
The most crucial problem with superstring theory is that
it is defined only perturbatively around consistent backgrounds,
and within such perturbative formulations,
it is known that 
the cosmic singularity is not resolved 
generally \cite{Liu:2002kb,Lawrence:2002aj,%
Horowitz:2002mw,Berkooz:2002je}.
In order to overcome this problem, one really needs to 
use a fully nonperturbative formulation.
In fact, there exist concrete proposals
for such a formulation using
supersymmetric matrix models \cite{IKKT,BFSS,DVV}.
These models can be obtained \emph{formally}\footnote{Note that
ten-dimensional ${\cal N}=1$ super Yang-Mills theory is not well-defined
at the quantum level due to gauge anomaly. The relationship between
the matrix models and the 10-dimensional super Yang-Mills theory actually
refers only to that of the classical action.} 
by dimensionally reducing
ten-dimensional ${\cal N}=1$ super Yang-Mills theory
to $d=0,1,2$ dimensions, respectively.
Based on these proposals, various issues of the early universe 
have been discussed \cite{Freedman:2004xg,Craps:2005wd,Li:2005sz,%
Das:2005vd,Chen:2005mga,She:2005mt,Martinec:2006ak,Ishino:2006nx,%
Matsuo:2008yd,Klammer:2009ku}.
As a closely related
direction, ref.~\cite{McFadden:2009fg} proposes a conformal 
field theory, which is holographically dual to inflationary models.
(See also ref.~\cite{Bzowski:2012ih} and references therein.)

The IIB matrix model \cite{IKKT} is one of the matrix model
proposals 
corresponding to the $d=0$ case mentioned above,\footnote{See 
ref.~\cite{Nishimura:2014vza}
for a review of recent developments in the IIB matrix model.} 
in which not only space
but also time should emerge dynamically from 
the matrix degrees of freedom.
This aspect of the IIB matrix model has also been discussed
intensively 
as ``emergent gravity'' \cite{Szabo:2006wx,%
Steinacker:2007dq,Steinacker:2008ri,Steinacker:2010rh,%
Polychronakos:2013fma,Yang:2006dk,Yang:2008fb,Yang:2013aia}
in noncommutative field theories,
which appear in the IIB matrix model
for a particular class of classical 
backgrounds \cite{Aoki:1999vr,Ambjorn:1999ts,Ambjorn:2000nb,Ambjorn:2000cs}.
(See also ref.~\cite{Hanada:2005vr} for a different proposal for 
emergence of curved space-time in the matrix model.)
Until quite recently,
the IIB matrix model 
was
studied after making a Wick rotation since
the partition function of the Euclidean matrix model 
obtained in this way was shown to be 
finite \cite{Krauth:1998xh,Austing:2001pk}.
In fact, a lot of efforts have been devoted
to 
identifying the matrix configurations that dominate
the partition function 
using various methods \cite{AIKKT,Hotta:1998en,Ambjorn:2000bf,%
Ambjorn:2000dx,Anagnostopoulos:2001yb,%
Nishimura:2000ds,Nishimura:2000wf,Nishimura:2001sx,Kawai:2002jk,Aoyama:2006rk,%
Imai:2003vr,Imai:2003jb,Imai:2003ja,%
Anagnostopoulos:2013xga,Nishimura:2011xy}.
However, the Euclidean matrix model is clearly not applicable to cosmology 
since it does not provide the real-time dynamics.
Moreover, it is known that 
the Wick rotation is more subtle in 
quantum gravity than in 
quantum field theory at the nonperturbative level (See, for instance,
refs.~\cite{Ambjorn:2005qt,Kawai:2011rj}).
Indeed a recent study based on the Gaussian expansion method
suggests that the space-time obtained dynamically in the Euclidean matrix
model does not seem to correspond to our four-dimensional 
space-time \cite{Nishimura:2011xy}.

Motivated by all these problems with the Euclidean IIB matrix model,
three of the authors (S.-W.K., J.N. and A.T.)
studied the Lorentzian version of the IIB matrix model
by Monte Carlo simulation for the first time \cite{KNT}.
Unlike the Euclidean version,
one has to introduce
infrared cutoffs in both spatial 
and temporal directions
in order to make the partition function finite.
However, it was found that these two cutoffs can be removed
in the large-$N$ limit in such a way that physical quantities scale.
The eigenvalue distribution of the matrix representing the time
extends in that limit,
and the dominant matrix configurations
have a very nontrivial structure, which enables 
us to naturally extract the time-evolution.
Quite surprisingly, it was found that
a phase transition occurs at some point in time,
and after that, only three out of nine spatial
directions start to expand.
This phase transition can be interpreted 
as the birth of our 3-dimensional universe
in superstring theory. It should be emphasized that 
the results seem to suggest that
the space-time
dimensionality is determined \emph{uniquely} by the nonperturbative
dynamics of superstrings unlike in perturbative string theory,
in which consistent backgrounds can have various space-time
dimensionality.

As another important property of the 
Lorentzian IIB matrix model,
it is expected that the classical approximation 
becomes valid at late times \cite{Kim:2011ts,Kim:2012mw}.
The reason for this is that
each term in the action 
has large contribution from the degrees of freedom at late times 
due to the expansion of the universe.
One can actually construct a simple solution representing an expanding
(3+1)-dimensional universe, which naturally 
solves the cosmological
constant problem \cite{Kim:2012mw}.
It has also been argued that local field theory emerges
from low-lying fluctuation modes around a solution representing a
commutative space-time \cite{Nishimura:2012rs}.
In fact, the classical equations of motion of the matrix model
have infinitely many solutions,
which is reminiscent of
the so-called landscape in superstring theory.
Unlike the situation with the landscape, however,
there is a definite criterion 
to pick up a particular solution describing the late-time behaviors
since we have a well-defined partition function.

Clearly 
it is important to extend 
the Monte Carlo studies in ref.~\cite{KNT}
to much longer time. 
For instance, it would be interesting 
to see whether the inflation and
the Big Bang occurs in this model as is generally believed 
in modern cosmology.
Moreover, if we can go further and reach the time region in which
the classical approximation is valid,
we should be able to determine the solution which actually describes the
late-time behaviors in the dominant configurations.
Once this has been done, we should be able to derive the effective field
theory below the scale where gravity decouples
by considering the fluctuations around the classical solution.
In particular, it would be interesting to see whether the Standard Model
particles appear at low energy, for instance, in a way speculated
in refs.~\cite{Chatzistavrakidis:2011gs,Aoki:2010gv,Nishimura:2013moa,%
Aoki:2014cya}.

To this end,
we develop a new method 
that enables us to
investigate a long time-evolution of the universe
in the Lorentzian IIB matrix model
by Monte Carlo simulation.
Note, in particular, 
that the time scale that one would hope to 
achieve is, at least,
a few orders of magnitude
larger than 
the typical time scale of the model.
If one attempts to study it directly, one would need 
a huge matrix size, 
which makes the calculation impractical.
In this paper we show that there exists
a ``renormalized theory'', which 
corresponds to a theory obtained from the original model
by integrating out the dynamical degrees of freedom 
at earlier times.
The renormalized theory has two extra parameters compared 
with the original model, which can be 
used to optimize the length of the time region that one can 
actually probe.
By simulating the renormalized theory with optimized parameters,
we can investigate
the late-time behaviors with much less dynamical degrees of freedom
than the direct approach would require.

In order to show how the method works, we consider a simplified
model, which can be 
obtained from the original model by
neglecting the coupling of fermionic matrices to the spatial bosonic
matrices. This approximation is expected to 
be valid at early times, where the space is not 
so large.
Then the Pfaffian that arises from integrating out fermionic matrices
can be expressed by
some power of the van der Monde determinant,
which is written explicitly in terms of the eigenvalues 
of the temporal matrix only,
and the simulation becomes as fast as the bosonic model.
The simplified model indeed
retains
the important properties of the
original model such as 
the spontaneous breaking of the rotational symmetry at some critical time.
Moreover, we find that 
the size of the universe grows exponentially 
after the critical time.
We apply the renormalization group method to the simplified
model and confirm the exponential expansion more clearly
with smaller matrices.

The rest of the paper is organized as follows.
In section \ref{sec:review} we review
some important properties of
the Lorentzian IIB matrix model.
In section \ref{sec:VDM} we define the simplified model,
and present results obtained by direct Monte Carlo studies.
In particular, we show that
the exponential expansion
is realized in this model 
after the spontaneous breaking of rotational symmetry.
In section \ref{sec:RGmethod} 
we describe the renormalization group method.
%
In section \ref{sec:scalingbehavior}
we apply the method to the simplified model
and show 
that it allows us to study the time-evolution 
much more efficiently.
Section \ref{sec:summary} is devoted to a summary and discussions.
In appendix \ref{sec:derivation} we derive the form of the
model suitable for Monte Carlo simulation.
In appendix \ref{sec:details} we give some details of our
Monte Carlo simulation.

\section{Brief review of the Lorentzian IIB matrix model}
\label{sec:review}

The IIB matrix model has an action \cite{IKKT}
\beqa
S &=& S_{\rm b} + S_{\rm f}  \ ,  
\label{IKKTaction}
\\
S_{{\rm b}} & = & 
-\frac{1}{4g^{2}}\mbox{Tr}\Bigl(
\left[A_{\mu},A_{\nu}\right]
\left[A^{\mu},A^{\nu}\right] \Bigr)
\ ,  
\label{S_b}
\\
S_{{\rm f}} & = & 
-\frac{1}{2g^{2}}\mbox{Tr}\Bigl(\Psi_{\alpha}
\left(\mathcal{C}\Gamma^{\mu}\right)_{\alpha\beta}
\left[A_{\mu},\Psi_{\beta}\right]\Bigr) \ ,
\label{S_f}
\eeqa
%
where the bosonic $N\times N$ matrices $A_{\mu}$ ($\mu=0,\cdots,9$)
and the fermionic ones $\Psi_{\alpha}$ ($\alpha=1,\cdots,16$)
are both traceless and Hermitian.
$\Gamma^{\mu}$ are
10-dimensional gamma-matrices after the Weyl projection
and $\mathcal{C}$ is the charge conjugation
matrix. 
Since the coupling constant $g$ can be absorbed 
by rescaling $A_{\mu}$ and $\Psi$ appropriately, it is merely
a scale parameter. 

The IIB matrix model is conjectured to be a nonperturbative
definition of superstring theory \cite{IKKT}.
There are various pieces of evidence for this conjecture.
First of all, the action 
(\ref{IKKTaction})
can be regarded as a matrix regularization of
the worldsheet action of type IIB superstring theory
in the Schild gauge \cite{IKKT}.\footnote{This does not imply 
that the matrix model is merely a formulation
for the ``first quantization'' of superstrings.
In fact, multiple worldsheets appear naturally
in the matrix model as block-diagonal configurations,
where each block represents the embedding of a single worldsheet
into the 10-dimensional target space.}
Secondly, D-branes in type IIB superstring theory can be
described in the matrix model, and the interaction between
them can be correctly reproduced \cite{IKKT}.
Thirdly, 
under a few reasonable assumptions,
the string field Hamiltonian for type IIB superstring theory
can be derived from Schwinger-Dyson equations for the Wilson loop
operators, which are identified as creation and annihilation operators
of strings \cite{Fukuma:1997en}.

In all these connections to type IIB superstring theory, 
the target space coordinates are identified with the eigenvalues
of the matrices $A_\mu$.
In particular, this identification is consistent with
the supersymmetry algebra of the model, in which the translation
that appears from the anti-commutator of supersymmetry generators
is identified with the shift symmetry
$A_\mu \mapsto A_\mu + \alpha_\mu {\bf 1}$
of the model,\footnote{Apparently, this 
symmetry is not consistent with the traceless condition on $A_\mu$.
For a more precise argument, one should consider a block diagonal
configuration and shift each block relatively \cite{IKKT}.}
where $\alpha_\mu \in {\bf R}$.
Also the fact that the model has extended ${\cal N}=2$ supersymmetry
in ten dimensions is consistent with the fact that the model 
actually includes gravity since it is known in field theory
that ${\cal N}=1$ supersymmetry is the maximal one that
can be achieved in ten dimensions without including gravity.

The partition function for the Lorentzian version of the IIB matrix 
model is proposed as \cite{KNT}
\beq
Z  =  \int dA \, d\Psi\, e^{i S} \ .
\label{part-lorentzian}
\eeq
Integrating out
the fermionic matrices, we obtain the Pfaffian
\beq
\mbox{\Pf}\,\mathcal{M}(A) 
= \int d\Psi\, e^{i S_{{\rm f}}} \ , 
\label{Pf-def}
\eeq
which is real.
Note that the bosonic action (\ref{S_b}) can be written as
\beqa
S_{{\rm b}}  &=&  
\frac{1}{4g^{2}}\mbox{Tr} ( F_{\mu\nu} F^{\mu\nu} )  
\label{Sb-F2}
\\
&=& 
\frac{1}{4g^{2}} \left\{  - 2 \, \mbox{Tr} ( F_{0i}  )^2 
+ \mbox{Tr} ( F_{ij}  )^2  \right\} \ , 
\label{F2-two-terms}
\eeqa
where we have defined Hermitian matrices
$F_{\mu\nu} =  i \, [ A_\mu , A_\nu ]$.
Hence the bosonic action is not positive semi-definite.

The partition function (\ref{part-lorentzian})
is not finite as it stands,
but it can be made finite
by introducing infrared cutoffs 
in both temporal and spatial directions as \cite{KNT}
\begin{eqnarray}
\frac{1}{N}\mbox{Tr}\left(A_{0}\right)^{2}  
& \leq &  \kappa \frac{1}{N}\mbox{Tr}\left(A_{i}\right)^{2}  \ ,
\label{kappa}\\
\frac{1}{N}\mbox{Tr}\left(A_{i}\right)^{2} & \leq & \Lambda^{2} \ .
\label{L}
\end{eqnarray}
It turned out that these two cutoffs can be
removed in the large-$N$ limit in such a way
that physical quantities scale.
The resulting theory thus obtained
has no parameter except one scale parameter.

After some manipulation
and rescaling of $A_\mu$
(See Appendix \ref{sec:derivation}), 
the partition function
can be rewritten as \cite{KNT}
\begin{align}
Z = & \int dA \,
{\rm Pf} {\cal M} (A) \,
\delta\left(
\frac{1}{N}\Tr (F_{\mu\nu}F^{\mu\nu})  \right)
\delta\left(\frac{1}{N}\Tr (A_i)^2 - L^2 \right)
\theta\left(\kappa L^2 - \frac{1}{N}\Tr (A_0)^2  \right)  \ ,
\label{our-model}
\end{align}
where $\theta(x)$ is the Heaviside step function
and $L$ is a scale parameter introduced for later convenience.
Since the Pfaffian ${\rm Pf} {\cal M}(A)$ is real
unlike in the Euclidean 
case \cite{Anagnostopoulos:2001yb,Nishimura:2000ds,%
Nishimura:2000wf,Anagnostopoulos:2013xga},
the model (\ref{our-model}) can be studied by Monte Carlo simulation
without the sign problem.\footnote{Strictly speaking, the Pfaffian
is not positive semi-definite. However, it turns out that the configurations
with positive Pfaffian dominate the partition function at large $N$.
Therefore, we may simply replace the Pfaffian by its absolute value
$|{\rm Pf} {\cal M}(A)|$ in actual simulation \cite{KNT}.}

Let us then discuss
how we can extract the time-evolution from a configuration
generated by simulating the
system (\ref{our-model}).
First we choose the SU($N$) basis in such a way that
the temporal matrix $A_0$ is diagonalized as
\beq
A_0= {\rm diag} (\alpha_1 , \cdots , \alpha_N) \ ,
\quad
\mbox{where~} 
\alpha_1 <  \cdots < \alpha_N \ .
\label{A0-gauge}
\eeq
In that basis, it turned out that the spatial matrices
$A_i$ has a band-diagonal structure with
off-diagonal elements $(A_i)_{ab}$ for $|a-b|\ge n$ being small
for some integer $n$.
Therefore, we may naturally consider
$n\times n$ block matrices
\beq
(\bar{A}_i) _{IJ} (t)
\equiv (A_i)_{ \nu +I, \nu +J } \ ,
\eeq
where $I,J = 1, \cdots ,n$ and $ \nu = 0 , 1 , \cdots , N-n$, 
as representing a state 
of the universe at the time
\beq
t=\frac{1}{n}\sum_{I=1}^{n}\alpha_{\nu+I} \ .
\eeq
For instance, the extent of space at time $t$ is defined by
\begin{equation}
R^{2}(t)=
\left\langle \frac{1}{n}\mbox{tr}\left(\bar{A}_{i}\left( t \right)\right)^{2} 
\right\rangle
\ ,
\label{def-R}
\end{equation}
where the trace here is taken over the $n\times n$ block.
In order to see the spontaneous breaking of SO(9) symmetry,
we define the ``moment of inertia tensor''
\begin{equation}
T_{ij}(t)=
\frac{1}{n}\mbox{tr}\left(\bar{A}_{i} (t)
\bar{A}_{j} (t) \right) \ ,
\label{T-def}
\end{equation}
which is represented by a real symmetric $9\times 9$ matrix.
We denote the real positive semi-definite eigenvalues of
$T_{ij}(t)$ as $\lambda_j(t)$ with the ordering
\beq
\lambda_1(t) > \lambda_2(t) >  \cdots > \lambda_9(t) \ . 
\eeq
If the SO(9) symmetry is not spontaneously broken, 
the expectation values $\langle \lambda_{i}(t) \rangle$ 
become equal in the large-$N$ (and large-$n$) limit.
We find that this is indeed the case at early times,
while at sufficiently late times, three of the eigenvalues become
considerably larger than the others, suggesting that 
the SO(9) symmetry
is spontaneously broken down to SO(3) after a critical time.

The necessity for introducing the cutoff (\ref{kappa})
in the temporal direction can be understood as follows.
Let us consider a situation in which the eigenvalues of $A_0$
are well separated from each other and estimate the effective
action for the eigenvalues perturbatively.
By fixing the gauge to (\ref{A0-gauge}), we rewrite the integration 
over $A_\mu$ as
\beqa
\label{gauge-fixing}
\int dA &=& \int dA_i \int\prod_{a=1}^{N}d\alpha_{a}\,
\Delta (\alpha)^2 \ , \\
\Delta (\alpha) &\equiv &
\prod_{a>b}^{N}
\left(\alpha_{a}-\alpha_{b}\right) \ ,
\label{A0diag}
\eeqa
where $\Delta(\alpha)$ is the van der Monde (VDM) determinant.
The action can be expanded as
\beqa
S_{{\rm b}} &=& -\frac{1}{4g^{2}} (\alpha _ a - \alpha _ b)^2 |(A_i)_{ab}|^2
+ \cdots \ , \\
S_{{\rm f}} & = & 
-\frac{1}{2g^{2}} (\Psi_{\alpha})_{ba} 
(\alpha _ a - \alpha _ b)
\left(\mathcal{C}\Gamma^{\mu}\right)_{\alpha\beta}
(\Psi_{\beta})_{ab}  + \cdots \ ,
\eeqa
where the omitted terms are subleading 
for large $|\alpha _ a - \alpha _ b|$.
Integrating out $A_i$ at the one-loop level, one obtains
$\Delta(\alpha)^{-18}$ neglecting the zero modes corresponding to 
diagonal elements.
Integrating out $\Psi_\alpha$ at the one-loop level, one obtains
$\Delta(\alpha)^{16}$ neglecting the zero modes.
Thus one finds that the $\Delta(\alpha)^2$
in (\ref{gauge-fixing})
is canceled exactly at the one-loop level,
which is actually a consequence of 
the supersymmetry \cite{IKKT} of the model (\ref{part-lorentzian}).
Due to this property, the eigenvalue distribution of $A_0$ 
extends to infinity even for finite $N$ if it were not for
the cutoff (\ref{kappa}).

\section{A simplified model and its properties}
\label{sec:VDM}

The argument given 
above
motivates us to generalize the Lorentzian IIB matrix model
to $(d+1)$-dimensional versions ($d=9,5,3$),
which can be obtained by dimensional reduction of $(d+1)$-dimensional 
$\mathcal{N}=1$ super Yang-Mills theory.
(The $d=9$ case corresponds to the Lorentzian IIB matrix model.)
In general, integration over $A_i$ gives
$\Delta(\alpha)^{-2d}$, while
integration over the fermionic matrices gives
$\Delta(\alpha)^{2(d-1)}$.
Hence the VDM determinant that appears as in (\ref{gauge-fixing})
is canceled exactly at the one-loop level
and there is no interaction among the eigenvalues of $A_0$ at the one-loop
level. In Monte Carlo simulations, the eigenvalue distribution 
indeed extends to infinity if one does not introduce 
the temporal cutoff $\kappa$ as in (\ref{kappa}).
If one omits fermions, one obtains an attractive force between the 
eigenvalues of $A_0$, and the eigenvalue distribution of $A_0$ has
a finite extent without any cutoff.

In fact,
the $d=5$ supersymmetric model
turns out to have
very similar properties as the original model.\footnote{Similarity 
of (5+1)-dimensional version and the (9+1)-dimensional
version is seen also in the Euclidean IIB matrix model.
It was found that SO($D$) symmetry 
of the $D$-dimensional model is broken
down to SO(3) symmetry for $D=10$ \cite{Nishimura:2011xy}
and $D=6$ \cite{10070883}, and
various properties associated with the SSB
turned out to be common to both models \cite{Nishimura:2011xy}.
}
In particular, the SO(5) rotational symmetry is broken spontaneously
down to SO(3) after a critical time.
The (5+1)-dimensional model contains bosonic matrices
$A_\mu$ ($\mu = 0, \cdots , 5$) and fermionic matrices 
$\Psi_\alpha$ and $\bar{\Psi}_\alpha$ ($\alpha = 1 , \cdots , 4 $).
The action for the fermionic matrices is given by
\beq
S_{{\rm f,6d}}  =  
-\frac{1}{2g^{2}}\mbox{Tr}\left( \bar{\Psi}_{\alpha}
\left(\Gamma^{\mu}\right)_{\alpha\beta}
\left[A_{\mu},\Psi_{\beta}\right]\right) \ ,
\label{S_f-6d}
\eeq
where $\Gamma^{\mu}$ are 6-dimensional gamma-matrices after 
the Weyl projection.
Integrating out the fermionic matrices, we obtain the determinant
\beq
\mbox{det}\,\mathcal{M}(A) 
= \int d\Psi\,d\bar{\Psi}\, e^{i S_{{\rm f,6d}}} \ , 
\label{det-def}
\eeq
which is real.
This model can be studied by Monte Carlo simulation
using the partition function (\ref{our-model}), 
where ${\rm Pf} {\cal M} (A)$ should be replaced by 
$\mbox{det}\,\mathcal{M}(A) $.

In performing 
Monte Carlo simulation of the model (\ref{our-model})
or its (5+1)-dimensional version,
the most time-consuming part comes from 
calculating the contribution from the fermions.
Here we consider a simplified model, which can be obtained
by replacing the Pfaffian (\ref{Pf-def})
or the determinant (\ref{det-def}) by $\Delta(\alpha)^{2(d-1)}$,
which we obtained as the leading contribution
when the separation $|\alpha _ a - \alpha _ b|$ of the eigenvalues
of $A_0$ is large.
Note that this amounts to neglecting the terms proportional to
the spatial matrices $A_{i}$ ($i=1,\cdots ,d$)
in the fermionic action (\ref{S_f}) or (\ref{S_f-6d}).
Therefore we expect that the simplified model captures
the qualitative behaviors of the original models at early times
before the expansion of space proceeds much.
Thus we arrive at the model
\begin{align}
Z_{\rm VDM} = & 
\int \prod_{a=1}^{N}d\alpha_{a} \prod_{i=1}^{d} dA_i \, 
\Delta (\alpha)^{2d} \, 
\delta\left(
\frac{1}{N}\tr (F_{\mu\nu}F^{\mu\nu})  \right)
 \nonumber \\
& \times
\delta\left(\frac{1}{N}\tr (A_i)^2 - L^2 \right)
\theta\left(\kappa L^2 - \frac{1}{N}\tr (A_0)^2  \right)  \ ,
\label{vdm-model}
\end{align}
where $A_0$ is given by (\ref{A0-gauge}).
This model, which we call the VDM model in what follows,
can be simulated as easily as the bosonic model.
Moreover, it shares important properties with the original 
supersymmetric models
such as the spontaneous symmetry breaking of SO($d$)
and expanding behavior of the three-dimensional space after 
the critical time.


\begin{figure}[t]
\centering\includegraphics[width=7.4cm]{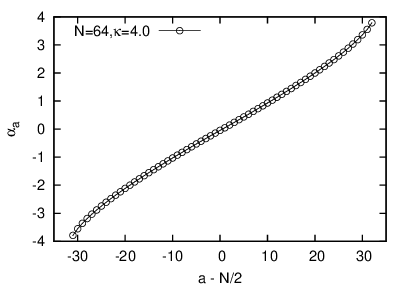}
\centering\includegraphics[width=7.4cm]{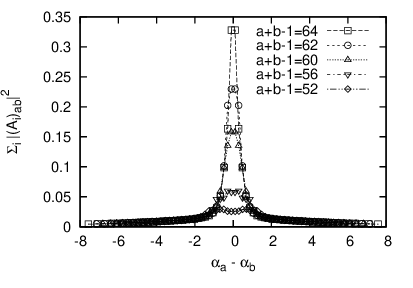}
\caption{(Left) The eigenvalues $\alpha_a$ of $A_0$
with the ordering (\ref{A0-gauge})
are plotted against its label $a$
for $N=64$ and $\kappa=4$. 
(Right) The magnitude of the off-diagonal elements of $A_i$ 
defined by the quantity $\sum_i |(A_i)_{ab}|^2$
is plotted against the time separation $\alpha_a - \alpha_b$.
}
\label{fig:eigen-offdiagonal}
\end{figure}



In this paper we study the VDM model in the $d=5$ case\footnote{The 
properties of the $d=5$ VDM model observed here are confirmed
also in the $d=9$ case. In particular, 
we observe the SSB from SO(9) to SO(3) at some critical time.
} 
for simplicity by Monte Carlo simulation.
(See appendix \ref{sec:details} for the details of our simulation.)
We set $L=1$ in (\ref{vdm-model})
without loss of generality since it 
only fixes the scale of the model.
%
In figure \ref{fig:eigen-offdiagonal} (Left) we plot
the eigenvalues $\alpha_a$ of $A_0$
with the ordering (\ref{A0-gauge})
against its label $a$ for $N=64$ and $\kappa=4$. 
Figure \ref{fig:eigen-offdiagonal} (Right) shows
the magnitude of the off-diagonal elements of $A_i$ against 
the time separation $\alpha_a - \alpha_b$. 
We find that the off-diagonal element
decreases rapidly as one goes away from the diagonal element.
Moreover, we observe a nice scaling behavior for 
sufficiently large $|\alpha_a - \alpha_b|$.
The region with small $|\alpha_a - \alpha_b|$ that does not scale
includes roughly 8 points. 
Based on this observation,
we choose the block size
to be $n=8$ in this paper.

\begin{figure}[t]
\centering\includegraphics[width=7.4cm]{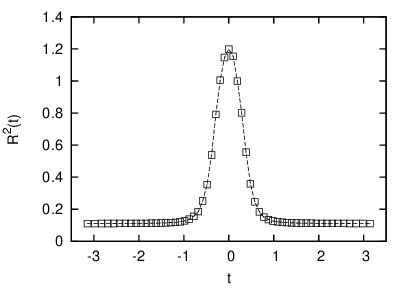}
\centering\includegraphics[width=7.4cm]{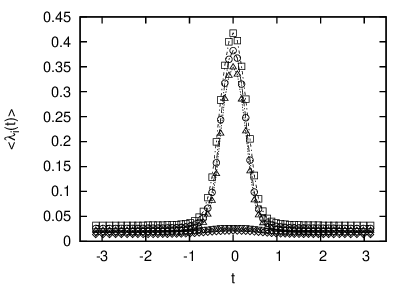}
\caption{(Left) The extent of space
$R^2(t)$ is plotted against
$t$ 
for $N=64$ and $\kappa=4$ with the block size $n=8$. 
(Right) The expectation values
$\langle \lambda_i(t) \rangle$
of the five eigenvalues of $T_{ij}(t)$ are plotted against 
$t$
for $N=64$ and $\kappa=4$ with the block size $n=8$. 
Three of them start to increase rapidly after a critical time.
From this behavior we define $t_{\rm c}$ as explained 
in the text.}
\label{fig:Rplot1}
\end{figure}

In figure \ref{fig:Rplot1} (Left)
we plot $R^{2}(t)$ against $t$ for $N=64$ and $\kappa=4.0$.
This plot shows that the space starts to expand 
at a critical time.
In figure \ref{fig:Rplot1} (Right) we plot the
expectation value $\langle \lambda_i(t) \rangle$
of the five eigenvalues of $T_{ij}\left(t\right)$,
which shows that the SSB from SO(5) to SO(3) occurs
at the critical time. 

The definition of the critical time $t_{\rm c}$ is 
ambiguous at finite $N$.
As a convenient choice we define it as follows.
Note first that
the appearance of a gap between 
$\langle \lambda_3(t) \rangle$ and 
$\langle \lambda_4(t) \rangle$
signals the SSB of SO(5) to SO(3).
Let us therefore
define the separation $d_j(t) = 
\langle \lambda_j(t) \rangle 
- \langle \lambda_{j+1}(t) \rangle$.
Then we find that the symmetric phase can be characterized by
$d_1(t) > d_2(t) > d_3(t) > d_4(t)$,
while in the broken phase we find $d_2(t) < d_3(t)$.
Therefore we define the critical time $t_{\rm c}$ by the largest value
of $t'$ such that
$d_1(t) > d_2(t) > d_3(t) > d_4(t)$ holds for $t \le t'$.
For instance, the critical time $t_{\rm c}$ obtained in this way
from figure \ref{fig:Rplot1} (Right) is $t_{\rm c} = -0.8813(2)$
and the extent of space at the critical time
is $R^2(t_{\rm c})=0.139(1)$.

\begin{figure}[t]
\centering\includegraphics[width=7.4cm]{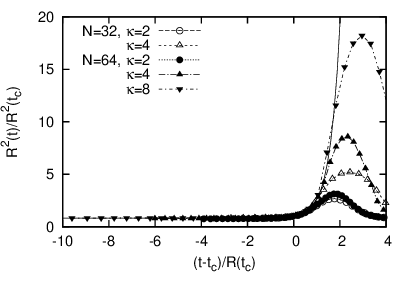}
\centering\includegraphics[width=7.4cm]{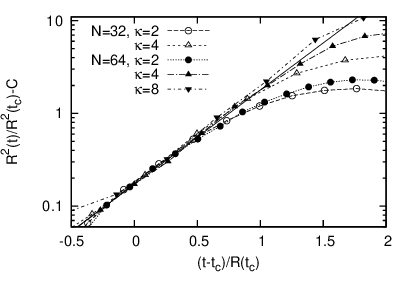}
\caption{(Left) The extent of space
$R^2(t)/R^2(t_{\rm c})$ is plotted against $(t-t_{\rm c})/R(t_{\rm c})$
for $N=32$ with $\kappa=2,4$ and for $N=64$ with $\kappa=2,4,8$.
The block size for the measurement is taken to be $n=8$ for all the cases. 
The solid line represents a fit of the data for 
$N=64$ and $\kappa=4$ to the behavior 
$R^2(t)/R^2(t_{\rm c}) = C + (1-C)\exp (b x)$ with $x=(t-t_{\rm c})/R(t_{\rm c})$,
which yields $C = 0.83(1)$ and $b=2.3(1)$.
%
(Right) The quantity $R^2(t)/R^2(t_{\rm c})-C$ is plotted 
against $(t-t_{\rm c})/R(t_{\rm c})$ in the log scale.
The constant $C$ is obtained from the exponential 
fit in the left panel,
which corresponds to the solid straight line in this plot.
%
}
\label{fig:Rplot2}
\end{figure}

In figure \ref{fig:Rplot2} (Left) we plot 
$R^2(t)/R^2(t_{\rm c})$ against $(t-t_{\rm c})/R(t_{\rm c})$
for various values of $\kappa$ and $N$, which reveals a
nice scaling property\footnote{In figure \ref{fig:Rplot2} alone,
we adjust the critical time $t_{\rm c}$
slightly from the point defined above
for each parameter set ($\kappa, N$)
in such a way that we optimize the scaling 
with the data for $\kappa=4$ and 
$N=64$, which are plotted without such adjustment.} 
of the function $R^2(t)$.
The shift in time is necessary since only
the difference $(t-t_{\rm c})$ is meaningful.
We also normalize all dimensionful quantities by $R(t_{\rm c})$,
which represents the size of the universe when it was born.
%
%
Interestingly, we observe that our data can be fitted well to 
\beq
\frac{R^2(t)}{R^2(t_{\rm c})} \equiv f(x) 
= C + \tilde{C} \exp (- b x)
\ , \quad
\mbox{where~}x =\frac{t - t_{\rm c}}{R(t_{\rm c})} \ .
\eeq
(We fix the second coefficient to be $\tilde{C} =1-C$ using
the constraint $f(0)=1$, which follows from the definition of $f(x)$.)
This is demonstrated in figure \ref{fig:Rplot2} (Right), 
where we plot $R^2(t)/R^2(t_{\rm c}) - C$ against 
$(t - t_{\rm c}) /R(t_{\rm c})$ in the logarithmic scale.
This exponential growth is reminiscent of the inflation, 
which is expected to have taken place in the early universe.
A similar behavior at early times can be seen also
in our preliminary results for the Lorentzian IIB matrix 
model \cite{Ito:2013qga}.

Let us discuss how we should take the large-$N$ limit.
For that we define the 
``lattice spacing'' $\epsilon$
and the ``time-extent'' $\Delta$ by
\beq
\epsilon  \equiv \frac{\delta t}{R(t_{\rm c})} \ , \quad \quad \quad
\Delta \equiv  \frac{t_{\rm p} - t_{\rm c}}{R(t_{\rm c})} \ ,
\eeq
where $\delta t$ is the mean separation of the eigenvalues of $A_0$
and $t_{\rm p}$ represents the time $t$ at which the extent of 
space $R(t)$ becomes maximum.
(In fact, $t_{\rm p}=0$ 
as one can see from figure \ref{fig:Rplot1}
due to the time reflection symmetry.)
Results for different $\kappa$ and $N$ correspond to 
different $\epsilon$ and $\Delta$.
As $\kappa$ is increased for a fixed $N$, the time-extent $\Delta$
increases and one can see late-time behaviors more.
However, the lattice spacing $\epsilon$ increases at the same time,
which results in deviations from the scaling behavior 
due to ``lattice artifacts''.
Therefore, one needs to increase $N$ as one increases $\kappa$
to see
the scaling behavior at later times.
The fact that the scaling behavior extends with increasing $N$
implies that the two cut-offs (\ref{kappa}) and (\ref{L})
can be removed in the large-$N$ limit.
Whether the time-extent $\Delta$ diverges in the large-$N$ limit
or not is an interesting dynamical question.
If it diverges, the $t>0$ region in figure \ref{fig:Rplot1}, for instance,
becomes invisible in the large-$N$ limit, hence 
there will be no Big Crunch.

In order to study the late-time behaviors, we need to increase 
the matrix size further. 
However, we notice from figure \ref{fig:Rplot2} (Left)
that the symmetric phase extends more than the broken phase
with increasing $N$. Due to this property of the model, it is not
efficient to study the late-time behaviors by just increasing 
the matrix size.

\section{Renormalization group method}
\label{sec:RGmethod}



\begin{figure}[t]
\centering\includegraphics[width=5cm]{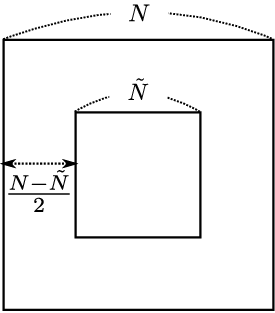}
\caption{
The basic idea of the renormalization group
in the Lorentzian matrix model. If we take the SU($N$) basis 
(\ref{A0-gauge}), the inner part of the matrices corresponds
to the late-time behaviors.}
\label{figure:renorm-matrix}
\end{figure}

In this section we propose a new method based on the idea of
renormalization group, which enables us to study the late-time
behaviors much more efficiently than in a direct approach.
Note first that
the late-time behaviors are described by the inner part of the matrices 
$A_\mu$ (See figure \ref{figure:renorm-matrix}.)
if we fix the gauge to (\ref{A0-gauge}).
The corresponding degrees of freedom are given by
$\tilde{N} \times \tilde{N}$ Hermitian matrices $\tilde{A}_\mu$,
which are defined by
\beq
(\tilde{A}_\mu)_{ab}
= (A_\mu)_{s + a , s + b } \ , \quad \quad s \equiv \frac{N-\tilde{N}}{2} \ ,
\eeq
where the indices $a$ and $b$ run from $1$ to $\tilde{N}$.
In principle, we can
derive the renormalized theory for $\tilde{A}_\mu$ by integrating out
the other degrees of freedom in the original matrices $A_\mu$.
Once we know the form of the renormalized theory, 
we can study the late-time behaviors efficiently 
by simulating the renormalized theory, which has much less degrees of
freedom than the original model.

In fact, the properties of the renormalized theory 
can be investigated
by simulating the original model written in terms of $A_\mu$
and measuring quantities written in terms of $\tilde{A}_\mu$ only.
In what follows, we put tildes on all the variables and parameters
of the renormalized theory.
For instance, corresponding to the cutoffs (\ref{kappa}), (\ref{L}),
we define $\tilde{\kappa}$ and $\tilde{L}$ 
for the renormalized theory by
\beq
\tilde{\kappa} \, \tilde{L}^{2}
 \equiv  
\left\langle \frac{1}{\tilde{N}} 
\mbox{Tr} ( \tilde{A}_{0} )^{2}  \right\rangle
 \ , \quad \quad
\tilde{L}^{2}
\equiv  
\left\langle \frac{1}{\tilde{N}}\mbox{Tr}
( \tilde{A}_i )^{2} \right\rangle
 \ ,
\label{tilde-kappaL}
\eeq
where the symbol $\langle \ \cdot \ \rangle$
refers to the VEV with respect to the original model
for the whole matrices.
Let us also define the quantities
\beq
 \tilde{B}  
\equiv  
\left\langle
\frac{1}{\tilde{N}}\mbox{Tr} (\tilde{F}_{ij})^{2} 
\right\rangle
 \ ,
\quad \quad
\tilde{E}  \equiv  
\left\langle \frac{2}{\tilde{N}}\mbox{Tr} (\tilde{F}_{0i})^{2}
\right\rangle  \ ,
\label{tilde-BE}
\eeq
where $\tilde{F}_{\mu\nu} =  i \, [ \tilde{A}_\mu , \tilde{A}_\nu ]$.
Note that
$ \left\langle 
\frac{1}{\tilde{N}} \mbox{Tr} (\tilde{F}_{\mu\nu}\tilde{F}^{\mu\nu})
\right\rangle
= \tilde{B} - \tilde{E}$ is not constrained to be zero unlike in the
original model (\ref{vdm-model}).

\begin{figure}[t]
\centering\includegraphics[width=10cm]{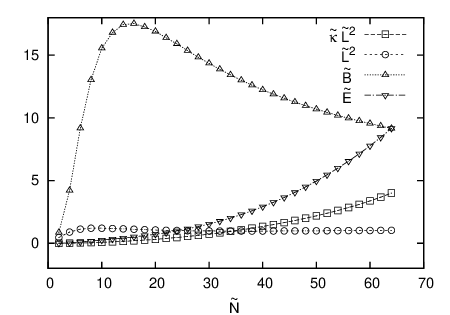}
\caption{The results for
$\tilde{\kappa}\tilde{L}^2$, $\tilde{L}^2$, $\tilde{B}$ and $\tilde{E}$ 
obtained in the original model
with $N=64$ and $\kappa=4$ are plotted against $\tilde{N}$.
}
\label{fig:param-flow}
\end{figure}

In figure \ref{fig:param-flow} the results for 
$\tilde{\kappa}$, $\tilde{L}$, $\tilde{B}$ and $\tilde{E}$ 
obtained from simulations of the original model
with $N=64$ and $\kappa=4$ are plotted against $\tilde{N}$ .
For $\tilde{N}=64$, which corresponds to the results for
the original model, we have 
$\tilde{\kappa}=4$, $\tilde{L}=1$, $\tilde{B}=\tilde{E}$
as it should.
Note also
that $\tilde{B}\neq \tilde{E}$ for $\tilde{N}<64$.

\begin{figure}[t]
\centering\includegraphics[width=10cm]{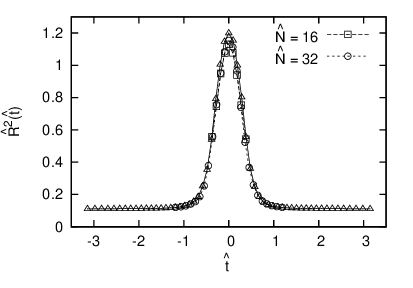}
\caption{The extent of space $R^2(t)$ obtained for the
effective theory with 
$N=16, 32$ 
is plotted.
They agree nicely with the data points (triangles) 
for the original model
with $N=64$ and $\kappa=4$, which are also plotted.
}
\label{fig:Rplot-renorm}
\end{figure}

Let us then consider an effective theory for 
the $\hat{N} \times \hat{N}$ Hermitian matrices $\hat{A}_\mu$.
(Here and henceforth, we put hats 
on all the variables
and parameters of the effective theory.)
We propose
\begin{align}
Z_{\rm eff} = & 
\int \prod_{a=1}^{\hat{N}} d\hat{\alpha}_{a} 
\prod_{i=1}^{d} d\hat{A}_i \, 
\Delta (\hat{\alpha})^{2d} \, 
\delta\left(
\frac{2}{\hat{N}}\tr (\hat{F}_{0i})^2 - \hat{E}  \right)
\delta\left(
\frac{1}{\hat{N}}\tr (\hat{F}_{ij})^2 - \hat{B}  \right)
 \nonumber \\
& \times
\delta\left(\frac{1}{\hat{N}}\tr (\hat{A}_i)^2 - \hat{L}^2 \right)
\theta\left(\hat{\kappa} \hat{L}^2 
- \frac{1}{\hat{N}}\tr (\hat{A}_0)^2  \right)  \ ,
\label{eff-theory}
\end{align}
where $\hat{A}_0$ is given by (\ref{A0-gauge}) with all the variables
replaced with the ones with hats.
Formally, the only difference from the original model
(\ref{vdm-model}) is that we constrain 
$\frac{2}{\hat{N}}\tr (\hat{F}_{0i})^2$ 
and $\frac{1}{\hat{N}}\tr (\hat{F}_{ij})^2$
separately to some values. 
We study the effective theory for various $\hat{N}$
with the parameters $\hat{\kappa}$, $\hat{L}$, 
$\hat{E}$ and $\hat{B}$ chosen
to be $\tilde{\kappa}$, $\tilde{L}$, $\tilde{B}$ and $\tilde{E}$
obtained for $\tilde{N} = \hat{N}$ in figure \ref{fig:param-flow}.
The results for $\hat{R}^2(\hat{t})$ obtained in this way
are plotted in figure~\ref{fig:Rplot-renorm}.
We find that they reproduce the late-time behaviors of the 
original model very well except the region near $\hat{t}=0$, which is
subject to finite ``volume'' effects anyway.
This 
demonstrates
that the effective theory (\ref{eff-theory}) indeed
captures the late-time behaviors of the original model with much smaller
matrix size. Note, in particular, that the symmetric phase, which is not
interesting to us, is reduced considerably compared with the original model.
In fact, the results for $\hat{N}=16$ do not have
a symmetric phase at all. We find it remarkable that the data points
agree with those for the original model even in this case.

In application of this method, we actually do not know 
in advance how to choose the parameters
$\hat{\kappa}$, $\hat{L}$, $\hat{E}$ and $\hat{B}$.
This does not spoil the usefulness of the approach at all.
In order to show it, we need to answer the question:
Can the effective theory (\ref{eff-theory}) 
for arbitrary values of $\hat{N}$, 
$\hat{\kappa}$, $\hat{L}$, $\hat{B}$ and $\hat{E}$
be regarded as 
a renormalized theory for $\tilde{A}_\mu$
in the above sense?
By counting the number of parameters in the theory,
one finds that the answer is generically ``yes''.
Let us consider the original VDM model with $N$ and $\kappa$.
Then we define (\ref{tilde-kappaL}) and (\ref{tilde-BE})
for submatrices of size $\tilde{N}$,
which we denote as 
$\tilde{\kappa}(\tilde{N} ; N , \kappa )$,
$\tilde{L}(\tilde{N} ; N , \kappa )$,
$\tilde{B}(\tilde{N} ; N , \kappa )$,
$\tilde{E}(\tilde{N} ; N , \kappa )$.
This specifies a renormalized theory.
We try to match it with the effective theory 
(\ref{eff-theory}) by setting 
$\tilde{N} = \hat{N}$.
In order to match $\hat{\kappa}$ and $\hat{L}$
with $\tilde{\kappa}(\tilde{N} ; N , \kappa )$
and $\tilde{L}(\tilde{N} ; N , \kappa )$,
we can always make a rescaling of $\hat{A}_0$ and $\hat{A}_i$ as
\beq
\tilde{A}_0 = 
\frac
{\sqrt{\tilde{\kappa}} \tilde{L}( \tilde{N} ; N , \kappa)} 
{\sqrt{\hat{\kappa}} \hat{L}}
\hat{A}_0
 \ , \quad 
\tilde{A}_i = 
\frac
{\tilde{L}( \tilde{N} ; N , \kappa )} 
{\hat{L}}
\hat{A}_i
 \ .
\label{rescalingA0Ai}
\eeq
$\hat{B}$ and $\hat{E}$ should be rescaled accordingly,
and we require that they should match 
(after the rescaling) 
with $\tilde{B}( \tilde{N} ; N , \kappa)$ 
and $\tilde{E}( \tilde{N} ; N , \kappa)$ as
\beqa
\tilde{B} ( \tilde{N} ; N , \kappa)  &=&  
\left(\frac
{\tilde{L}( \tilde{N} ; N , \kappa)}
{\hat{L}}
\right)^4
\hat{B}
 \ , \nonumber \\
\tilde{E}( \tilde{N} ; N , \kappa ) &=&  
\left( \frac
{\sqrt{\tilde{\kappa}}\tilde{L} ( \tilde{N} ; N , \kappa )}
{\sqrt{\hat{\kappa}} \hat{L}}
\frac{\tilde{L}( \tilde{N} ; N , \kappa )}{\hat{L}}
\right)^2
\hat{E}
  \ .
\label{BE-matching}
\eeqa 
Since we have two arbitrary parameters $N$ and $\kappa$ 
at our disposal,
we can always choose them to satisfy the two conditions
in (\ref{BE-matching}).
(Strictly speaking, since $N$ can take only integer values,
the above statement holds as good approximation 
for sufficiently large $N$.)

Thus we have shown that
the effective theory (\ref{eff-theory}) 
for arbitrary values of $\hat{N}$, $\hat{\kappa}$, $\hat{L}$, 
$\hat{B}$ and $\hat{E}$
can be regarded as a renormalized theory of the original model
for the submatrices corresponding to late times.
For this to work, it was necessary to 
make the rescaling (\ref{rescalingA0Ai}).
This implies that when one makes a plot like the one
in figure \ref{fig:Rplot2} (Left)
for the effective theory (\ref{eff-theory}),
one should note
that the quantity for the $x$-axis 
is related to the corresponding
quantity in the renormalized theory through
\beq
\frac{\tilde{t}-\tilde{t}_{\rm cr}}
{\tilde{R}(\tilde{t}_{\rm cr})}
=\frac{\hat{t}-\hat{t}_{\rm cr}}{z \hat{R}(\hat{t}_{\rm cr})} \ ,
\label{t-tilde-t}
\eeq
where the time-rescaling parameter $z$ is given by
\beq
z = \frac{\sqrt{\hat{\kappa}}}
{\sqrt{\tilde{\kappa} ( \tilde{N} ; N , \kappa )} }  \ .
\label{def-a}
\eeq
Therefore we need to plot our results against
the right-hand side of (\ref{t-tilde-t}).
Since we do not know $\tilde{\kappa} ( \tilde{N} ; N , \kappa )$
in (\ref{def-a}) a priori, we 
determine the parameter $z$ in such a way that
the results for the model (\ref{eff-theory}) 
scale with the results for the original model
at earlier times. In the next section we will show that this is 
indeed possible, and the method 
enables us to study the late-time behaviors
of the original model in a much more efficient way.

\section{Scaling behaviors in the effective theory}
\label{sec:scalingbehavior}

In this section we show how the renormalization group method
works by simulating the model (\ref{eff-theory}).
From now on, we omit the hats on all
the variables and the parameters of the model (\ref{eff-theory}).
We study various values of $B$ and $E$ with $N=32$, 
$\kappa=4$ and $L=1$ fixed.
From a simulation of the original model with $\kappa=4$ and $N=32$, 
we get $B=7.5$ and $E=7.6$.
The incomplete cancellation $E-B \sim 0.1$ is due to 
numerical artifacts from finiteness of $\gamma_{C}$
in (\ref{Vpot-def}).

\begin{figure}[t]
\centering\includegraphics[width=7.4cm]{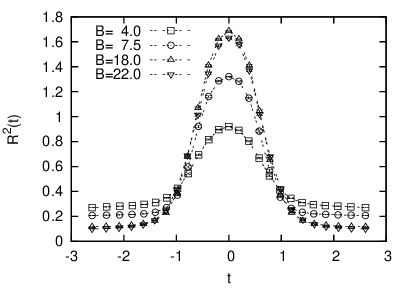}
\centering\includegraphics[width=7.4cm]{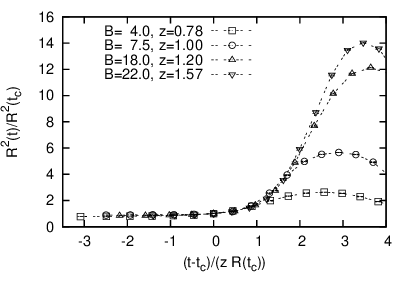}
\caption{(Left) The extent of space $R^2(t)$ 
is plotted against $t$ for various $B$ 
(including $B=7.5$, which corresponds to the original model)
with fixed $N=32$, $\kappa=4$ and $E=7.6$.
(Right) The same data are plotted in physical units.
The time-rescaling parameter $z$ is chosen
in such a way that 
the data scale with the results for $B=7.5$
(corresponding to the original model), for which we use $z=1$.
}
\label{fig:e-fixed-1}
\end{figure}

In figure \ref{fig:e-fixed-1} (Left)
we show our results for the model (\ref{eff-theory}) 
with various $B$ 
(including $B=7.5$, which corresponds to the original model)
for fixed $E=7.6$.
In figure \ref{fig:e-fixed-1} (Right)
we plot the same quantity in physical units.
We have introduced the time-rescaling parameter $z$,
which is set to $z=1$ for $B=7.5$, which corresponds 
to the original model, and otherwise it is chosen in such a way that
the results scale with the results for $B=7.5$ at earlier times.
Indeed we observe good scaling behavior
as anticipated from our arguments in the previous section.
We also find that the number of data points in the symmetric phase
($t < t_{\rm c}$) decreases as $B$ increases.
This means that we can use the matrix degrees of freedom
more efficiently for the more interesting broken phase.
On the other hand, we find that the lattice spacing $\epsilon$
increases slowly as we increase $B$.
We have to make sure that the lattice artifacts are kept under control
when we increase $B$.

\begin{figure}[t]
\centering\includegraphics[width=7.4cm]{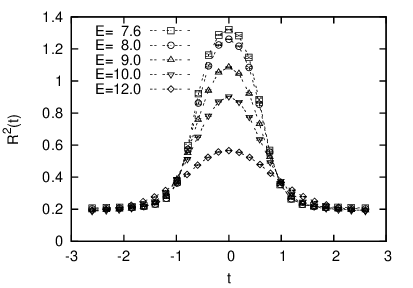}
\centering\includegraphics[width=7.4cm]{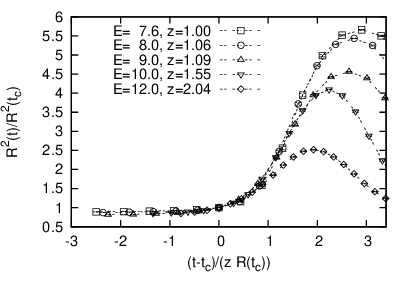}
\caption{(Left) The extent of space $R^2(t)$ 
is plotted against $t$ for various $E$ 
(including $E=7.6$, which corresponds to the original model)
with fixed $N=32$, $\kappa=4$ and $B=7.5$.
(Right) The same data are plotted in physical units.
The time-rescaling parameter $z$ is chosen
in such a way that
the data scale with the results for $E=7.6$
(corresponding to the original model), for which we use $z=1$.
}
\label{fig:b-fixed}
\end{figure}

Figure \ref{fig:b-fixed} (Left) shows our results for
the model (\ref{eff-theory}) with various $E$
(including $E=7.6$, which corresponds to the original model)
for fixed $B=7.5$.
In figure \ref{fig:b-fixed} (Right) we plot the same quantity
in physical units.
The time-rescaling parameter $z$
is set to $z=1$ for $E=7.6$, which corresponds 
to the original model, and otherwise it is chosen in such a way that
the results scale with the results for $E=7.6$ at earlier times.
We observe good scaling behavior
as anticipated from our arguments in the previous section.
We also find that the number of data points in the symmetric phase
($t < t_{\rm c}$) decreases as $E$ increases.
On the other hand, we find that the lattice spacing $\epsilon$
and hence the time-extent $\Delta$
decrease rapidly as we increase $E$.



\begin{figure}[t]
\centering\includegraphics[width=8.5cm]{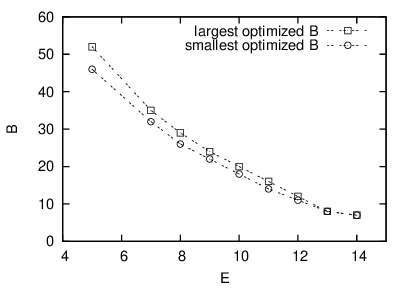}
\caption{The largest $B$ (squares) and the smallest $B$ (circles)
that correspond to the case with only one
data point in the symmetric phase
for a fixed $E$ is plotted against $E$.
}
\label{fig:b-e}
\end{figure}

The above results suggest a simple strategy for optimizing $B$ and $E$.
First we increase $B$ from the value for the original model
until the lattice spacing $\epsilon$ becomes a bit too large.
Then we can increase $E$ a little in order to 
make the lattice spacing $\epsilon$ sufficiently small.
If necessary, we repeat this procedure a few times until
the number of data points in the symmetric phase becomes sufficiently 
small. This way we can increase the time-extent $\Delta$
keeping the lattice spacing sufficiently small with the same matrix size.
In figure \ref{fig:b-e} the region
between the two curves in the $B$-$E$ plane correspond to the case
with only one data point in the symmetric phase. 
Within this region, the lattice spacing $\epsilon$ decreases
as one increases $E$.

\begin{figure}[t]
\centering\includegraphics[width=7.4cm]{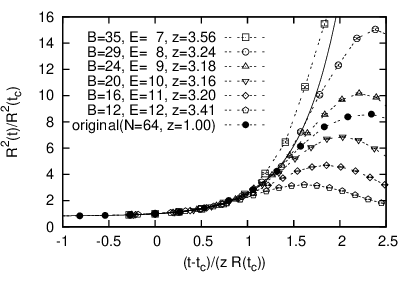}
\centering\includegraphics[width=7.4cm]{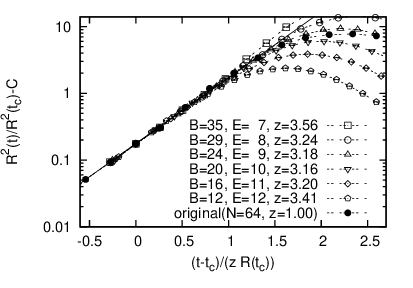}
\caption{(Left) The extent of space $R^2(t)$ is
plotted in physical units
for the parameter points $(B,E)$ on the upper curve
in figure \ref{fig:b-e}.
The time-rescaling parameter $z$ is fixed by referring
to the results for the original model with $N=64$ and $\kappa=4$.
(Right) $R^2(t)/R^2(t_{\rm c})-C$ is plotted in the log scale
against $(t-t_{\rm c})/(z R(t_{\rm c}))$,
where $C$ is determined by fitting the 
data in the left panel for $N=32$, 
$B=29$, $E=8$
to an exponential behavior 
$R^2(t)/R^2(t_{\rm c}) = C + (1-C)\exp (b x)$ with 
$x=(t-t_{\rm c})/(z R(t_{\rm c}))$,
which yields $C = 0.82(1)$ and $b=2.28(4)$.
}
\label{fig:optimized}
\end{figure}

Figure \ref{fig:optimized} (Left) shows the results
for the parameter points $(B,E)$ 
on the upper curve in figure \ref{fig:b-e}. 
In order to determine the time-rescaling parameter $z$,
we use the results for the original model with $N=64$ and $\kappa=4$
as a reference. 
The lattice spacing $\epsilon$ is larger for larger $B$ and smaller $E$.
We find that it becomes too large for $B=35$ and $E=7$
judging from the deviation from the scaling behavior.
The scaling region extends until one reaches
$B=29$ and $E=8$. This gives the maximum time-extent $\Delta$ 
that one can probe using the renormalization group method with $N=32$.

In figure \ref{fig:optimized} (Right) 
we plot $R^2(t)/R^2(t_{\rm c})-C$ against 
$(t-t_{\rm c})/(z R(t_{\rm c}))$,
where $C$ is determined by fitting the data in the left panel
to an exponential behavior 
$R^2(t)/R^2(t_{\rm c}) = C + (1-C)\exp (b x)$ 
with $x=(t-t_{\rm c})/(z R(t_{\rm c}))$.
We observe a clear straight line behavior providing
strong evidence for the exponential expansion of 
the early universe in the VDM model.



\section{Summary and discussions}
\label{sec:summary}

In this paper we have developed a new method
for studying the Lorentzian IIB matrix model for 
a long time period based on the idea of
the renormalization group.
The method is tested in a simplified model,
which captures the behaviors of the supersymmetric model
at early times.
We were able to confirm the exponential expansion of the
space observed in the simplified model with much smaller
matrix size using the new method.

On the conceptual side, we consider it interesting that
the idea of the renormalization group works in the Lorentzian 
matrix model.
The renormalization group has been applied to matrix models
some time ago by 
refs.~\cite{Brezin:1992yc,%
Higuchi:1993nq,Higuchi:1994rv,Nishigaki:1996np,Higuchi:1996np}
and more recently by refs.~\cite{Kawamoto:2012ng,Kawamoto:2013laa}.
In particular, ref.~\cite{Kawamoto:2013laa} studies
a Yang-Mills two-matrix model as a simplified model of the
Euclidean IIB matrix model.
In these papers some elements of the matrices were integrated out
explicitly to obtain a renormalized theory for matrices of
smaller size.
A crucial difference from these works is that
we have a notion of time in the Lorentzian matrix model.
This allows us to
consider a renormalized theory for the submatrices
representing the degrees of freedom at later times.

The effective theory for the submatrices representing the
later time behaviors contains two extra parameters.
We have shown how one can optimize them to probe the late-time
behaviors most efficiently.
The time-rescaling parameter denoted by $z$ has to be determined
by requiring that the results should scale at earlier times
with the results obtained for the original model.
This procedure becomes more complicated if one applies the
present method to the original supersymmetric model
since the fermionic action (\ref{S_f}) or (\ref{S_f-6d})
contains two terms; one of them being proportional to $A_0$
and the other being proportional to $A_i$.
The necessity to rescale $A_0$ and $A_i$ differently
as in (\ref{rescalingA0Ai}) requires us to 
make the coefficient
of the term proportional to $A_i$ 
in the fermionic action of
the effective theory, a new unknown parameter.
This new parameter can be fixed by probing 
the scaling behavior.
Despite this complication, we consider that
the renormalization group method is useful in
extracting the late-time behaviors in the 
Lorentzian IIB matrix model.
In particular, from the viewpoint of cosmology,
we consider it important to 
confirm the exponential expansion 
observed in our preliminary results 
for the Lorentzian IIB matrix model
reported in ref.~\cite{Ito:2013qga},
and to see whether it turns into
a power-law expansion at later times as suggested there.
We hope to report on these issues in future publications.

\section*{Acknowledgment}

Computation was carried out on PC clusters at KEK
and supercomputers SR16000 at YITP, Kyoto University 
and FX10 at University of Tokyo.
The work of 
Y.~I.\ is supported by Grant-in-Aid for 
JSPS
fellows.
The work of S.~-W.~K.\
is supported by the National Research Foundation
of Korea (NRF) Grant funded by the Korean Government
(MEST 2005-0049409 and NRF-2009-352-C00015).
The work of J.~N.\ and A.~T.\ is supported
by Grant-in-Aid for Scientific
Research
(No.\ 20540286, 24540264, and 23244057)
from JSPS.

\appendix

\section{Derivation of eq.~(\ref{our-model})}
\label{sec:derivation}

In this appendix we give a more in-depth derivation 
of eq.~(\ref{our-model}) than the one given in
the original paper \cite{KNT}.

Let us note first
that the integrand of the partition function (\ref{part-lorentzian})
involves a phase factor $e^{iS_{\rm b}}$.
As is commonly done in integrating oscillating functions,
we introduce the convergence factor
$e^{- \epsilon |S_{\rm b}|}$ and take the $\epsilon \rightarrow 0$ limit
after the integration.

The partition function can then be rewritten as
\begin{align}
Z &= \int dA \int_0^{\Lambda^2} dr \,
\delta\left(
\frac{1}{N}\mbox{Tr}\left(A_{i}\right)^{2} - r \right)
\theta\left(\kappa r - \frac{1}{N}\Tr (A_0)^2  \right) 
 e^{i S_{\rm b}-\epsilon |S_{\rm b}| }
\, {\rm Pf} {\cal M} \  ,
\end{align}
where $\kappa$ and $\Lambda$ are the cutoff parameters
introduced in (\ref{kappa}) and (\ref{L}), respectively.
Rescaling the variables $A_\mu \rightarrow r^{1/2} A_\mu$
in the integrand,
we get
\begin{align}
Z = & \int dA \,
{\rm Pf} {\cal M} (A) \,
f(S_{\rm b}) \,
\delta\left(\frac{1}{N}\Tr (A_i)^2 - 1 \right)
\theta\left(\kappa  - \frac{1}{N}\Tr (A_0)^2  \right)  \ ,
\label{our-model2}
\end{align}
where the function $f(S_{\rm b})$ is defined by
\beq
f(S_{\rm b}) \equiv
\int_0^{\Lambda^2}  dr  \, r^{9(N^2-1)-1}
 e^{r^2 (i S_{\rm b}-\epsilon |S_{\rm b}|) }
 \ .
\label{integrate-r}
\eeq
Note that $f(S_{\rm b})$
is a complex-valued function with the property 
$f(-S_{\rm b}) = f(S_{\rm b})^{*}$.
For $|S_{\rm b}| \ll \frac{1}{\Lambda^4}$,
the function can be well approximated by 
\beq
f(S_{\rm b}) \approx 
\frac{1}{9(N^2-1)} (\Lambda^2)^{9(N^2-1)} \ .
\label{f-Sb-small}
\eeq
For $|S_{\rm b}| \gtrsim \frac{1}{\Lambda^4}$,
the phase of the integrand in (\ref{integrate-r})
starts to oscillate violently in the region 
$r \gtrsim 1/\sqrt{|S_{\rm b}|}$,
and hence the integral decreases rapidly in magnitude
for increasing $|S_{\rm b}|$.
In particular, 
the asymptotic behavior 
of $f(S_{\rm b})$
for $|S_{\rm b}| \gg \frac{1}{\Lambda^4}$ can be 
estimated as
\beq
\frac{f(S_{\rm b})}{f(0)} 
=
\Gamma \left( \frac{9}{2}(N^2-1) +1 \right)
\,
\left( 
\frac{1}{\Lambda^4 |S_{\rm b}|} \right)^{\frac{9}{2}(N^2-1)} 
+ {\cal O}(e^{-\epsilon \Lambda^4 |S_{\rm b}| })
\eeq
by deforming the integration contour in (\ref{integrate-r}).
Recalling eq.~(\ref{Sb-F2}),
the condition $|S_{\rm b}| \ll \frac{1}{\Lambda^4}$
for (\ref{f-Sb-small}) can be rewritten as
\beq
\left| \frac{1}{N}\Tr (F_{\mu\nu}F^{\mu\nu})\right|  \ll 
\frac{4 g^2}{N \Lambda^4} \ .
\label{F2-condition}
\eeq
Therefore, assuming that the right-hand side 
$\frac{4 g^2}{N \Lambda^4}$ 
of (\ref{F2-condition})
becomes small at large $N$,
we may make a replacement
\beq
f(S_{\rm b}) 
\Longrightarrow
\delta\left(\frac{1}{N}\Tr (F_{\mu\nu}F^{\mu\nu})  \right)
\eeq
up to a normalization constant.
Rescaling the variables $A_\mu \rightarrow A_\mu/L$,
we arrive at eq.~(\ref{our-model}).
Within the above approximation, the parameter $L$ simply sets 
the scale of the model,
and we may use $L=1$ without loss of generality. 
This is also the case with the VDM model (\ref{vdm-model}).

\section{Details of Monte Carlo simulation}
\label{sec:details}

In this appendix we explain how we actually deal with the simplified 
model (\ref{vdm-model}) in Monte Carlo simulation.
Generalization to the effective theory (\ref{eff-theory}) is
straightforward.

First we replace the delta functions and the step function
in (\ref{vdm-model}) by Gaussian potentials as
\beqa
V_{\rm pot} & =&  
\frac{1}{2}\gamma_{C}
\left(\frac{1}{N}\mbox{Tr}\left(F_{\mu\nu}\right)^{2}\right)^{2}
 +\frac{1}{2}\gamma_{L}\left(\frac{1}{N}\mbox{Tr}\left(A_{i}\right)^{2}
-L^{2}\right)^{2} \nonumber \\
&~& +\frac{1}{2}\gamma_{\kappa}
\left(\frac{1}{N}\mbox{Tr}\left(A_{0}\right)^{2}
- \kappa L^{2} \right)^{2}
\theta \left(\frac{1}{N}\mbox{Tr}\left(A_{0}\right)^{2}
- \kappa L^{2} \right)  \ ,
\label{Vpot-def}
\eeqa
where the coefficients $\gamma_{C}$, $\gamma_{L}$, $\gamma_{\kappa}$
should be taken large enough to fix each observable
to the specified value. 
In actual simulation we have used $\gamma_{C}\sim1\times N^{2}$
and $\gamma_{L}=\gamma_{\kappa}\sim 100\times N^{2}$.

Another important issue concerns the spontaneous breaking
of the shift symmetry 
$A_0 \mapsto A_0 + \alpha {\bf 1}$.
For instance, when we try to calculate the expectation value
$R^{2}(t)$
defined in (\ref{def-R}),
the peak of the quantity measured for each configuration
fluctuates considerably.
This simply reflects the ambiguity in choosing the origin 
of the time coordinate, and we should fix it somehow before
taking the ensemble average.
Here we fix it by introducing
a potential\footnote{Strictly speaking, the shift symmetry is broken by
the traceless condition on $A_0$ and the cutoff (\ref{kappa}).
However, this breaking is not strong enough to solve the problem.}
\begin{eqnarray}
V_{{\rm sym}}
&=& \frac{1}{2}\gamma_{\rm sym}
\left(\frac{1}{N}
\left[\mbox{Tr}\,(A_{i})^{2}\right]_{{\rm left}}-
\frac{1}{N}\left[\mbox{Tr}\, (A_{i})^{2}\right]_{{\rm right}}\right)^{2} \ , \\
\left[ \mbox{Tr}\, (A_{i})^{2}\right]_{{\rm left}} 
& = & \sum_{i=1}^{d}\sum_{a+b<N+1}\left|(A_{i})_{ab}\right|^{2} \ ,\\
\left[\mbox{Tr}\,  (A_{i})^{2}\right]_{{\rm right}} 
& = & \sum_{i=1}^{d}\sum_{a+b>N+1}\left|(A_{i})_{ab}\right|^{2} \ ,
\label{V-sym-def}
\end{eqnarray}
where the coefficient is typically taken to be $\gamma_{{\rm sym}}\sim 100$.
We have checked that the results do not alter within error bars
for larger values of $\gamma_{{\rm sym}}$.

To summarize, the model we study by Monte Carlo simulation is given by
\begin{eqnarray}
Z_{\rm VDM} & = & 
\int\prod_{a=1}^{N}d\alpha_{a}\prod_{i=1}^{d}dA_{i}\, e^{-S_{\rm VDM}}\ ,
\nonumber \\
S_{\rm VDM} & = & -2d\log\Delta\left(\alpha\right)+ V_{\rm pot} + 
V_{\rm sym} \ .
\label{Z-VDM-def}
\end{eqnarray}

We apply the Hybrid Monte Carlo (HMC) method to simulate 
the model (\ref{Z-VDM-def}).
First we rewrite the model 
by introducing auxiliary variables $p_a$ and 
$(X_{i})_{ab}$ ($a,b=1,\cdots ,N$) with the action
\begin{equation}
S_{\rm HMC}  =  \frac{1}{2} \sum_a (p_a)^{2}
+\frac{1}{2} \mbox{Tr}(X_{i})^{2}
+S_{{\rm VDM}} [\alpha, A] \ .
\label{HMC-def}
\end{equation}
Here $p_a$ are real variables with the constraint
$\sum_a p_a = 0$, whereas
$X_{i}$ are traceless Hermitian matrices.
We update all the variables in the model (\ref{HMC-def})
as follows.
First we regard $p_a$ as the conjugate momenta of $\alpha_a$
and $X_{i}$ as the conjugate momenta of $A_i$.
Then we regard $S_{\rm HMC}$ in (\ref{HMC-def})
as the Hamiltonian $H$ and solve the classical equations of motion
obtained as the Hamilton equations
\begin{eqnarray}
\frac{d\alpha_a}{d\tau}  =  
\frac{\partial H}{\partial p_a}=p_a \ , &\quad&
\frac{d p_a}{d\tau}  =  
-\frac{\partial H}{\partial\alpha_a}=
-\frac{\partial S_{{\rm VDM}}}{\partial\alpha_a} \ ,
\nonumber
\\
\frac{dA_{i}}{d\tau}  =  
\frac{\partial H}{\partial X_{i}}=X_{i}^{*} \ ,
&\quad &
\frac{dX_{i}}{d\tau}  =  
-\frac{\partial H}{\partial A_{i}}=
-\frac{\partial S_{{\rm VDM}}}{\partial A_{i}} \ ,
\label{Ham-eq}
\end{eqnarray}
for some fictitious time $\tau$.
This part of the algorithm is called the Molecular Dynamics.
In solving the Hamilton equations (\ref{Ham-eq}) numerically,
we discretize them using the so-called leap-frog discretization,
which maintains reversibility with respect to $\tau$.
Starting from the previous configuration at $\tau =0$, we obtain
a new configuration at $\tau=\tau_{{\rm f}}$
by solving (\ref{Ham-eq}) with the step size $\Delta\tau$
so that $\tau_{{\rm f}}=N_{\tau}\cdot\Delta\tau$, where
$N_{\tau}$ is the number of steps.
We accept the new configuration with the probability
$\min (1, \exp (- \Delta S_{\rm HMC}))$, where
$\Delta S_{\rm HMC} \equiv S_{\rm HMC} (\tau_{{\rm f}})
- S_{\rm HMC} (0)$,
based on the idea of 
the Metropolis algorithm
to satisfy the detailed balance.
The crucial point here is that $S_{\rm HMC}$ is 
nothing but the Hamiltonian $H$,
which is preserved in the classical dynamics
if the equations (\ref{Ham-eq}) are solved exactly.
In fact, $\Delta S_{\rm HMC}$ is non-zero due to the 
discretization, but it is a small quantity of order $(\Delta \tau)^2$.
Therefore, one can move around efficiently
in the configuration space.

Since the auxiliary 
variables $p_a$ and $(X_{i})_{ab}$ appear only
as the Gaussian terms in (\ref{HMC-def}),
we can update them independently by 
using normalized Gaussian random numbers.
This procedure of refreshing the conjugate momenta
should be done each time we start a Molecular Dynamics procedure.
Thus the HMC algorithm as applied to our system
can be described as follows.
\begin{enumerate}
\item Generate initial configurations of $p_a(0)$
and $X_i(0)$
with the Gaussian distribution 
$e^{-\frac{1}{2}{\rm Tr}(X_i)^{2}}$
and 
$e^{-\frac{1}{2} \sum_a (p_a)^2}$, respectively.
\item Evolve the fields $p_a (\tau)$, $X_i (\tau)$,
$\alpha_a (\tau)$ and $A_i(\tau)$
for fictitious time $\tau_{{\rm f}}$ 
according to the discretized Molecular Dynamics.
\item Accept the obtained configuration 
of $\alpha_a (\tau_{{\rm f}})$ and $A_i(\tau_{{\rm f}})$
with the probability $\min (1,e^{-\Delta H})$, 
where $\Delta H=H(\tau_{{\rm f}})-H(0)$.
\end{enumerate}

The HMC algorithm involves two 
parameters $\Delta\tau$ and $\tau_{\rm f}$,
which can be optimized.
(See, for instance, appendix B of ref.~\cite{Ambjorn:2000dx}
for more details.)
For fixed $\tau_{\rm f}$, we have to choose $\Delta\tau$ so that
$\Delta\tau\times\left(\mbox{acceptance rate}\right)$
is maximized. Typically this is achieved for acceptance
rate of 50$\sim$60\%.
Then $\tau_{{\rm f}}$ can be optimized to minimize
the autocorrelation time in units of one step in the Molecular Dynamics.


\end{document}